\documentclass[aps,prl,groupedaddress,twocolumn]{revtex4}

\usepackage{graphicx}

\begin{document}

\title{Ultrafast Optical-Pump Terahertz-Probe Spectroscopy of the Carrier Relaxation and Recombination Dynamics in Epitaxial Graphene}
\author{Paul A. George, Jared Strait, Jahan Dawlaty, Shriram Shivaraman, Mvs Chandrashekhar, Farhan Rana, Michael G. Spencer}
\affiliation{School of Electrical and Computer Engineering, Cornell University, Ithaca, NY 14853}

\begin{abstract}
The ultrafast relaxation and recombination dynamics of photogenerated electrons and holes in epitaxial graphene are studied using optical-pump Terahertz-probe spectroscopy. The conductivity in graphene at Terahertz frequencies depends on the carrier concentration as well as the carrier distribution in energy. Time-resolved studies of the conductivity can therefore be used to probe the dynamics associated with carrier intraband relaxation and interband recombination. We report the electron-hole recombination times in epitaxial graphene for the first time. Our results show that carrier cooling occurs on sub-picosecond time scales and that interband recombination times are carrier density dependent.
\date{\today}
\end{abstract}


\maketitle

Graphene is a 2D lattice of carbon atoms arranged in a honeycomb crystal structure with a zero (or near-zero) bandgap and a linear energy-momentum dispersion relation for both electrons and holes \cite{dresselhaus_book,castroneto_bgnd}. The unique electronic and optical properties of graphene make it a promising material for the development of high-speed electron devices, including field-effect transistors, pn-diodes, Terahertz oscillators, and electronic and optical sensors \cite{castroneto_bgnd,novoselov_bgnd,zhang_bgnd,liang_bgnd,berger_bgnd,rana_bgnd}. The realization of graphene-based devices requires understanding the non-equilibrium carrier dynamics as well as the rate at which electron-hole recombination occurs.

Measurements of the ultrafast intraband relaxation dynamics of photogenerated electrons and holes in epitaxial graphene using both degenerate \cite{dawlaty_pupr} and non-degenerate \cite{norris_pupr} optical-pump optical-probe spectroscopy have been previously reported. Similar measurements for exfoliated graphene mono- and multi-layers have also been carried out \cite{heinz_pupr}. These measurements were sensitive to the interband conductivity of graphene and probed the time evolution of the carrier occupation at specific energies in the bands. Consequently, they were not able to directly measure the time scales associated with carrier recombination. At room temperature, the optical response of graphene in the THz frequency range is described by the intraband conductivity -- the free carrier response -- which depends not only on the total carrier concentration but also on the carrier distribution in energy \cite{dawlaty_optx}. Therefore, THz radiation can be used to study the carrier relaxation and recombination dynamics in graphene. In this paper, we present results obtained from optical-pump THz-probe spectroscopy of epitaxial graphene in which the time-dependent conductivity of graphene that has been excited with an optical pump pulse is probed with a few-cycle THz pulse. We observe cooling of the photogenerated carrier distribution as well as electron-hole recombination in graphene in real time. Our results indicate that the recombination times in graphene depend on the carrier density and material disorder.

The epitaxial graphene samples used in this work were grown on the carbon face of semi-insulating 6$H$-SiC wafers using techniques that have been reported previously \cite{deheer_epitax}. As discussed in \cite{dawlaty_pupr,dawlaty_optx}, X-ray photoemission, Raman, and optical/IR/THz transmission spectroscopy were used to characterize each sample to determine the number of carbon atom layers and the carrier momentum relaxation time. In this work, sample B consists of $\sim$12 carbon atom layers with a momentum relaxation time $\tau \sim 20~\rm{fs}$ and sample C consists of $\sim$19 layers with $\tau \sim 4~\rm{fs}$ \cite{dawlaty_optx}. Raman spectroscopy (at 488 nm) of the samples revealed a $G$ peak close to 1580 cm$^{-1}$ and a $D$ peak close to 1350 cm$^{-1}$. The $D$ peak is forbidden in perfect graphene layers since it requires an elastic scattering process, which is made possible by disorder, to satisfy momentum conservation \cite{faugeras_disorder,ferrari_disorder}. The presence of the D peak therefore indicates the presence of disorder in the samples. The average ratio of the intensity of the Raman $G$ and $D$ peaks, $I_{\rm G}/I_{\rm D}$, was 17 for sample B and 2 for sample C. This indicates significantly more disorder in the latter sample \cite{faugeras_disorder,ferrari_disorder}. Results from optical-pump optical-probe spectroscopy of sample B were reported earlier \cite{dawlaty_pupr}.

Ultrafast 90 fs pulses from a 81 MHz Ti:Sapphire laser with a center wavelength of 780 nm (1.6 eV) were used to simultaneously pump the THz Time-Domain Spectrometer (THz-TDS) and the epitaxial graphene samples. The THz-TDS used a photoconductive emitter and an electro-optic detector  \cite{grisch_tx,zhang_rx}, and generated few-cycle THz pulses with a 0.3-3.0 THz bandwidth and a power SNR near $10^6$. The epitaxial graphene samples were placed at the focus of a $90^\circ$ off-axis parabolic mirror and excited with 1 nJ - 16 nJ optical pulses focused to FWHM spot sizes of ~350 $\mu$m - 400 $\mu$m. The time-dependent transmission of the THz probe pulse through the optically excited sample was measured by delaying the optical pump pulse with respect to the THz probe. The pump and probe signals were mechanically chopped at 256 Hz and 179.2 Hz, respectively, and the change in the transmission was measured using a lock-in amplifier referenced to the sum of these frequencies.

The complex amplitude transmission of the THz pulse through the graphene layers and the SiC substrate (normalized to the transmission through the SiC substrate) is,

\begin{equation}
\label{eq_tx}
t = \frac{1}{1 + N \eta_{o} \, \sigma_{\rm intra}  / (1+n_{\rm SiC}) }
\end{equation}
\noindent $N$ is the number of carbon atom layers, $\eta_{o}$ is the free-space impedance, and $n_{\rm SiC}$ is the refractive index of SiC and equals $\sim$2.55. The transmission is related to the complex intraband conductivity of graphene, $\sigma_{\rm intra}$, which is given by \cite{dawlaty_optx},
\begin{equation}
\label{eq_intracond}
\sigma_{\rm intra} = i \frac{e^2 / \pi\hbar^2 }{\omega + i/\tau} \int_0^\infty \left[ f \left( \epsilon - \epsilon_{\rm Fc} , T \right) + f \left( \epsilon + \epsilon_{\rm Fv} , T \right) \right] d \epsilon
\end{equation}
\noindent In this expression, $\tau$ is the momentum relaxation time, $f(\epsilon-\epsilon_{\rm F},T)$ is the Fermi-Dirac distribution with carrier temperature $T$, and $\epsilon_{\rm Fc}$ and $\epsilon_{\rm Fv}$ are the Fermi energies of conduction band and valence band electrons, respectively. The electron and hole densities in graphene are,

\begin{equation}
\label{eq_stats}
\left( \begin{array}{r}
n \\
p
\end{array} \right) =  \frac{2}{\pi v^2 \hbar^2} \int_0^\infty f \left( \begin{array}{r}  \epsilon - \epsilon_{\rm Fc} \\  \epsilon + \epsilon_{\rm Fv} \end{array}, T \right) \epsilon \ d \epsilon
\end{equation}

\noindent Equations \ref{eq_intracond} and \ref{eq_stats} show that, unlike in semiconductors with parabolic energy bands, the intraband conductivity of graphene depends on the carrier distribution in addition to the carrier density. In graphene, a hotter carrier distribution corresponds to a smaller intraband conductivity than a cooler distribution with the same carrier density. In intrinsic photoexcited graphene layers, $n=p$ and $ \epsilon_{\rm Fc} = -  \epsilon_{\rm Fv}$. In epitaxial graphene, the first one or two carbon atom layers adjacent to the SiC substrate are expected to have a large carrier density \cite{deheer_epitax, norris_pupr}.  However, the change in the THz intraband conductivity, as measured in our experiments, is dominated by the large number of intrinsic layers. The contribution of the doped layers to our measurements has therefore been ignored in the discussion that follows.

\begin{figure}[tbp]
\centering
\includegraphics[width=7.5cm]{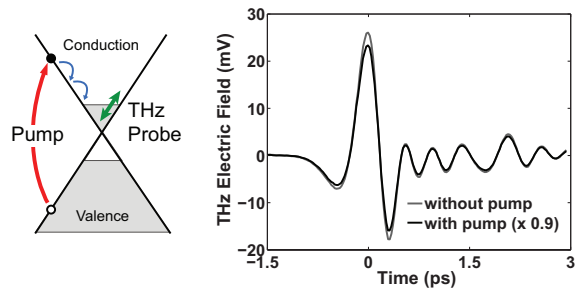}
\caption{Measured THz pulses transmitted through the epitaxial graphene sample B without (grey) and with (black, scaled) an optical pump pulse preceding the peak of the THz pulse by 1 ps. The the intraband conductivity of both samples B and C is almost entirely real and is nearly dispersionless at low THz frequencies. Consequently, the amplitude of the peak of the THz probe pulse is altered in the presence of the optical excitation but the pulse shape is not distorted nor is it shifted in time. Ringing after the pulse is due to absorption from water vapor.}
\label{fig_OPTP_time}
\end{figure}

\begin{figure}[hbp]
\centering
\includegraphics[width=7.03cm]{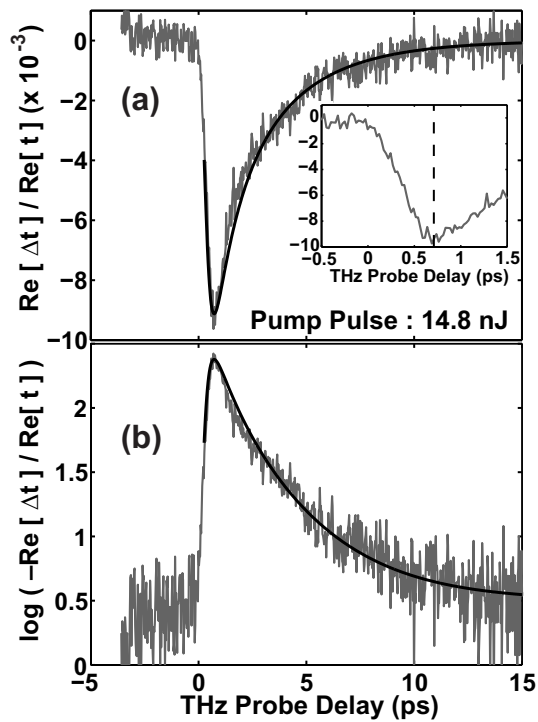}
\caption{(a) The measured change in the real part of the complex amplitude transmission (grey) and the theoretical fit (black) for pump pulse energy of 14.8 nJ (sample B). The transmission decreases rapidly until 0.75 ps - 1 ps and then increases slowly from 1 ps - 15 ps. The inlay shows the initial rise in the transmission, which occurs on a timescale longer than the experimental resolution of 130 fs. (b) The same data plotted on a logarithmic scale. The slow increase in the transmission does not follow an exponential curve.}
\label{fig_dt_exp}
\end{figure}

Figure \ref{fig_OPTP_time} shows the measured THz pulse transmitted through graphene sample B without and with (scaled for clarity) optical excitation for a probe delay of 1 ps. In this work, we measured the change in the transmission of the peak of the THz probe pulse caused by optical excitation. Because of the small momentum relaxation time of samples B and C, the real part of the intraband conductivity is much larger than the imaginary part and is nearly dispersionless in the low THz frequency range. In addition, the first order contribution from the change in the imaginary part of the conductivity to the change in the pulse transmission is exactly zero at the peak of the THz pulse. Consequently, near the peak, optical excitation primarily affects the amplitude of the THz probe pulse and does not distort the pulse shape or shift the pulse in time. Figure \ref{fig_dt_exp}(a) displays the experimentally measured time-dependent change in the real part of the normalized transmission of sample B resulting from excitation by a 14.8 nJ optical pulse. The time evolution of the transmission exhibits two distinct features: (1) an initial rapid decrease until 0.75 ps - 1 ps (left of dashed vertical line in inlay) followed by (2) a slow increase from 1 ps - 15 ps. The initial rapid decrease, shown in the inlay in Fig.~\ref{fig_dt_exp}, is within the (estimated) 130 fs temporal resolution of our measurements and is consistent with an increase in conductivity due to cooling of the thermalized carrier distribution after photoexcitation. The slow increase in the transmission is due to the decrease in conductivity resulting from electron-hole recombination and does not follow a simple exponential dependence (Fig.~\ref{fig_dt_exp}(b)). This is indicative of density-dependent carrier recombination times.

\begin{figure}[tbp]
\centering
\includegraphics[width=7.5cm]{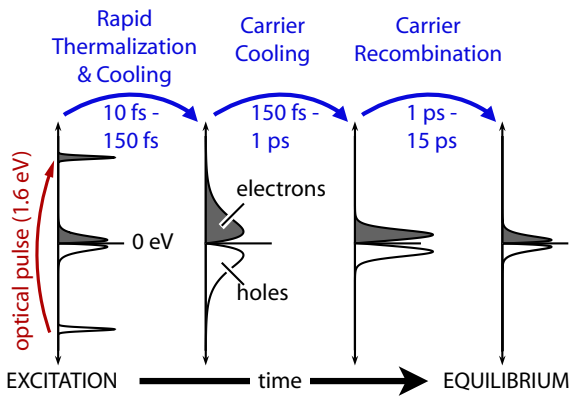}
\caption{A schematic of the likely processes by which optically-excited, non-equilibrium electron and hole distributions approach equilibrium. After excitation, the distribution rapidly thermalizes and cools. The hot thermally distributed carriers are then cooled further due to intraband phonon scattering. Finally, electrons and holes recombine until the equilibrium distribution is restored.}
\label{fig_recomb}
\end{figure}

\begin{figure}[tbp]
\centering
\includegraphics[width=7.03cm]{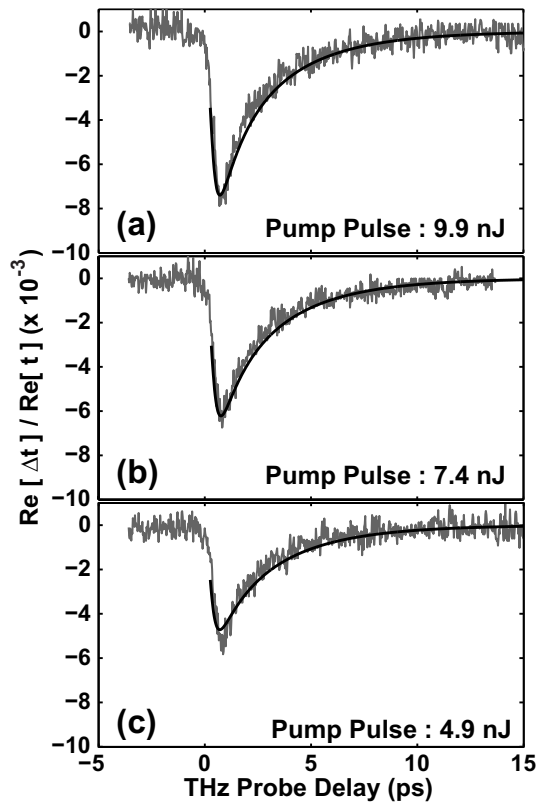}
\caption{The theoretically predicted and experimentally measured changes in the real part of the transmission for optical-pump pulse energies of (a) 9.9 nJ, (b) 7.4 nJ, and (c) 4.9 nJ (sample B). The model captures both the fast decrease and the slow increase of the transmission for different pump pulse energies.}
\label{fig_dt_theory}
\end{figure}

The likely processes by which non-equilibrium photogenerated electron and hole distributions return to equilibrium are shown pictorially in Fig.~\ref{fig_recomb}. Immediately following optical excitation, electrons (holes) are in a non-thermal distribution with a large population near 0.8 (-0.8) eV from the band edge. Within 10 fs - 150 fs, this distribution thermalizes and cools to form a hot Fermi-Dirac distribution \cite{dawlaty_pupr}. Intraband phonon scattering further cools the thermalized carriers between 150 fs and 1 ps. Beyond $\sim$1 ps, electron-hole recombination is the dominant processes affecting the intraband conductivity and the THz transmission. Intraband relaxation times in the 0.4 ps - 1.7 ps range were measured by the authors previously using optical-pump and optical-probe spectroscopy of epitaxial graphene \cite{dawlaty_pupr}. The intraband relaxation times measured in this work are slightly different. This small difference can be attributed to the fact that in \cite{dawlaty_pupr} the time-dependent optical transmission depended on the carrier occupation at a specific energy in the high energy tail of the distribution, whereas in the present case the time-dependent THz transmission is sensitive to the changes in the distribution at all energies.

Simple coupled rate equations can be used to model the carrier dynamics after optical excitation and subsequent thermalization. These equations model the cooling of the carrier temperature $T$ towards the equilibrium temperature with a phenomenological time constant $\tau_{\rm c}$ and the decay of the carrier population due to recombination. The mechanisms responsible for electron-hole recombination in graphene could include plasmon emission, phonon emission, and Auger scattering. In general, recombination rates due to these processes are expected to have non-trivial carrier density dependencies \cite{rana_bgnd,rana_auger,hwang_plasmons,harrison_book}. Additionally, carrier generation rates cannot be ignored in the analysis \cite{rana_auger}. We assume that the net recombination rate $R(n)$ can be expressed as a quadratic function of $n$, $R(n) = B (n^2 - n_{\rm eq}^2)$, where it is assumed that $n=p$ and $n_{\rm eq}$ is the carrier density at thermal equilibrium. $B$ is a fitting parameter. From the time-dependent carrier temperature $T$ and density $n$, the conductivity $\sigma_{\rm intra}$ and the THz transmission can be determined.

\begin{figure}[tbp]
\centering
\includegraphics[width=7.03cm]{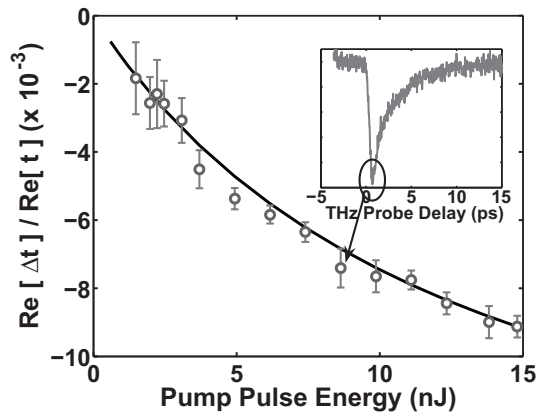}
\caption{The measured and the predicted peak change in the real part of the complex THz transmission for different pulse energies (sample B).}
\label{fig_dt_max}
\end{figure}

The agreement between the model (black) and the measured data (grey) for the THz probe transmission transient is also shown in Fig.~\ref{fig_dt_exp}. For a pump pulse energy of 14.8 nJ, the photoexcited carrier density was estimated to be $\sim 5 \times 10^{11}~\rm{cm}^{-2}$. To best fit the measured data, the carrier temperature immediately after thermalization was assumed to be 1250 K, and the values of $\tau_{\rm c}$ and $B$ were found to be 200 fs and $1.8~\rm{cm}^{2}~s^{-1}$, respectively. To explore the carrier density dependence of the measured transients, experiments were conducted with optical pump pulse energies varying from 1 nJ to 15 nJ. The model also exhibits an excellent fit to the data for different pump pulse energies, as shown in Fig.~\ref{fig_dt_theory} for pump pulse energies of 9.9, 7.4, and 4.9 nJ. In the model, the initial photoexcited carrier density and the initial carrier temperature were scaled linearly with the pump pulse energy \cite{dawlaty_pupr}. It is important to emphasize here that the model captures the initial decrease in the transmission, the peak change in the transmission, and the slow increase in the transmission for different pulse energies without changing the values of $B$ or $\tau_{\rm c}$. Figure \ref{fig_dt_max} shows good agreement between the measured and the calculated peak change (absolute minimum, see inlay) in the THz transmission for pump pulse energies varying from 1 nJ to 15 nJ. The extracted value of $\tau_{\rm c}$ agrees well with the calculated and measured inelastic carrier-phonon intraband scattering times in graphene \cite{sarma_transport} and in carbon nanotubes \cite{moos_cnt,perebeinos_cnt}. Using the extracted value of $B$, the calculated recombination rates for carrier densities in the $10^{11}-10^{12}~\rm{cm}^{-2}$ range are approximately 1.5-2.0 times larger than the theoretically predicted recombination rates due to Auger scattering in graphene \cite{rana_auger}. This difference can be attributed to additional recombination mechanisms, such as plasmon and phonon emission. It should be mentioned that the net recombination rate $R(n)$ can have a carrier density dependence different from the assumed quadratic expression. For example, in \cite{rana_auger} it has been shown that the Auger recombination rate in graphene can have a cubic dependence on the carrier density. The coefficient $B$ can also be carrier density dependent \cite{rana_auger}. However, for the values of the carrier density relevant to our experiments the calculated auger recombination rates have an almost quadratic dependence on the carrier density \cite{rana_auger}. In addition, within the resolution of our experiments, a constant value of $B$ adequately fits the data.

In \cite{dawlaty_pupr} and \cite{dawlaty_optx} it was shown that the intraband energy and momentum relaxation rates in epitaxial graphene depend on material disorder. Disorder could also affect the interband recombination rates since disorder can modify the density of states \cite{sarma_pupr} and/or provide the additional momentum necessary to satisfy the conservation rules in recombination processes \cite{grein_pupr,take_pupr}. Optical-pump THz-probe spectroscopy was performed on epitaxial graphene sample C which, as discussed earlier, is significantly more disordered than sample B. The value of $B$ for sample C was obtained using the methods discussed above and equals $3.1~\rm{cm}^{2}~s^{-1}$. The carrier recombination rate is approximately twice as fast in sample C than in sample B for the same carrier density.

In conclusion, we studied the ultrafast relaxation and recombination dynamics of photogenerated carriers in epitaxial graphene using optical-pump THz-probe spectroscopy. We measured the interband electron-hole recombination rates in graphene for the first time. Our measurements show that the recombination times in graphene depend on the carrier density.

\begin{acknowledgements}
The authors acknowledge support from the National Science Foundation, the Air Force Office of Scientific Research (Contract No. FA9550-07-1-0332, monitor Dr. Donald Silversmith), and the Cornell Material Science and Engineering Center (CCMR) program of the National Science Foundation (Cooperative Agreement No. 0520404). The authors would also like to thank Dr. Daniel Grischkowsky and Dr. Adam Bingham from Oklahoma State University for help in the design of the THz emitters used in this work.
\end{acknowledgements}


\begin{thebibliography}{99}

\bibitem{dresselhaus_book} R. Saito, G. Dresselhaus and M. S. Dresselhaus, \textit{Physical Properties of Carbon Nanotubes}, Imperial College Press, London, UK (1999).

\bibitem{castroneto_bgnd} A. H. Castro Neto, F. Guinea, N. M. R. Peres, K. S. Novoselov and A. K. Geim, arXiv:con-mat/0709.1163 (2007).

\bibitem{novoselov_bgnd} K. S. Novoselov, A. K. Geim, S. V. Morozov, D. Jian, M. I. Katsnelson, I. V. Grigorieva, S. V. Dubonos and A. A. Firsvo, Nature \textbf{438}, 197 (2005).

\bibitem{zhang_bgnd} Y. Zhang, Y. Tan, H. L. Stormer and P. Kim, Nature \textbf{438}, 201 (2005).

\bibitem{liang_bgnd} G. Liang, N. Neophytou, D. E. Nikonov and M. S. Lundstrom, IEEE Trans. Elec. Dev. \textbf{54}, 657 (2007).

\bibitem{berger_bgnd} C. Berger, Z. Song, X. Li, X. Wu, N. Brown, C. Naud, D. Mayou, T. Li, J. Hass, A. N. Marchenkov, E. H. Conrad, P. N. Frist and W. A. de Heer, Science \textbf{213}, 1191 (2006).

\bibitem{rana_bgnd} F. Rana, IEEE Trans. Nanotechnology \textbf{7}, 91 (2008).

\bibitem{dawlaty_pupr} J. M. Dawlaty, S. Shivaraman, M. Chandrashekhar, F. Rana and M. G. Spencer, Appl. Phys. Lett. \textbf{92}, 042116 (2008).

\bibitem{norris_pupr} D. Sun, Z. Wu, C. Divin, X. Li, C. Berger, W. A. de Heer, P. First and T. Norris, arXiv:con-mat/0803.2883 (2008).

\bibitem{heinz_pupr} D. Song, K. F. Mak, Y. Wu, C. H. Lui, M. Sfeir, S. Rosenblatt, H. Yan, J. Maultzsch and T. Heinz, Bul. of Am. Phys. Soc. \textbf{53}, L29.7 (2008).

\bibitem{dawlaty_optx} J. M. Dawlaty, S. Shivaraman, J. Strait, P. George, M. Chandrashekhar, F. Rana and M. G. Spencer, arXiv:con-mat/0801.3302 (2008).

\bibitem{deheer_epitax} W. A. de Heer, C. Berger, X. Wu, P. N. First, E. H. Conrad, X. Li. T. Li, M. Sprinkle, J. Hass, M. Sadowski, M. Potemski and G. Martinez, arXiv:con-mat/0704.0285 (2007).

\bibitem{faugeras_disorder} C. Faugeras, A. Nerriere, M. Potemski, A. Mahmood, E. Dujardin, C. Berger and W. A. de Heer, arXiv:con-mat/0709.2538 (2000).

\bibitem{ferrari_disorder} A. C. Ferrari and J. Robertson, Phys. Rev. B \textbf{61}, 14095 (2000).

\bibitem{grisch_tx} A. L. Bingham and D. Grischkowsky, Appl. Phys. Lett. \textbf{90}, 091105 (2007).

\bibitem{zhang_rx} Q. Chen, M. Tani and X. C. Zhang, J. Opt. Soc. Am. B \textbf{18}, 823 (2001).

\bibitem{sarma_transport} W. K. Tse, E. H. Hwang, and S. Das Sarma, arXiv:con-mat/0806.0436 (2008).

\bibitem{moos_cnt} T. Hertel and G. Moos, Phys. Rev. Lett. \textbf{84}, 5002 (2000).

\bibitem{perebeinos_cnt} V. Perebeinos, J. Tersoff and P. Avouris, Phys. Rev. Lett. \textbf{94}, 086802 (2005).

\bibitem{hwang_plasmons} E. H. Hwang and S. Das Sarma, Phys. Rev. B \textbf{75}, 205418 (2007).

\bibitem{rana_auger} F. Rana, Phys. Rev. B \textbf{76}, 155431 (2007).

\bibitem{harrison_book} P. Harrison, \textit{Quantum Wells, Wires, and Dots}, John Wiley and Sons, NY (USA) (2005).

\bibitem{sarma_pupr} B. Y. Hu, E. H. Hwang, S. D. Sarma, arXiv:con-mat/0805.2148 (2008).

\bibitem{grein_pupr} C. H. Grein, H. Ehrenreich, J. Appl. Phys., \textbf{93}, 1075 (2003).

\bibitem{take_pupr} M. Takeshima, Phys. Rev. B., \textbf{23}, 771 (1981).



\end{thebibliography}
\end{document}